\documentclass[times, twoside]{StyleArXiv}
\usepackage{booktabs}
\usepackage{multicol,multirow}
\usepackage{orcidlink}
\usepackage{subcaption}
\usepackage{cleveref}
\crefname{figure}{Fig}{Figs}
\crefname{table}{Tab}{Tabs}
\crefname{equation}{eq.}{eqs.}

\leadauthor{Siddhad} % Please give the surname of the lead author for the running footer

\begin{document}

\title{Awake at the Wheel: Enhancing Automotive Safety through EEG-Based Fatigue Detection}
\shorttitle{Awake at the Wheel: Enhancing Automotive Safety through EEG-Based Fatigue Detection}

\author[1,\Letter]{Gourav~Siddhad~\orcidlink{0000-0001-5883-3863}}
\author[2]{Sayantan~Dey~\orcidlink{0009-0009-4210-8435}}
\author[1]{Partha~Pratim~Roy~\orcidlink{0000-0002-5735-5254}}
\author[2]{Masakazu~Iwamura~\orcidlink{0000-0003-2508-2869}}
\affil[1]{Department of Computer Science and Engineering, Indian Institute of Technology, Roorkee, Uttarakhand, 247667, India}
\affil[2]{Department of Core Informatics, Graduate School of Informatics, Osaka Metropolitan University, Sakai, Osaka, 599-8531, Japan}

\maketitle

%%%%%%%%%%%%%%%%%%%%%%%%%%%%%%%%%%%%%%%%%%%%%%%%%%
%%%%%%%%%%%%%%%%%%%%%%%%%%%%%%%%%%%%%%%%%%%%%%%%%%

\begin{abstract}
Driver fatigue detection is increasingly recognized as critical for enhancing road safety. This study introduces a method for detecting driver fatigue using the SEED-VIG dataset, a well-established benchmark in EEG-based vigilance analysis. By employing advanced pattern recognition technologies, including machine learning and deep neural networks, EEG signals are meticulously analyzed to discern patterns indicative of fatigue. This methodology combines feature extraction with a classification framework to improve the accuracy of fatigue detection. The proposed NLMDA-Net reached an impressive accuracy of 83.71\% in detecting fatigue from EEG signals by incorporating two novel attention modules designed specifically for EEG signals, the channel and depth attention modules. NLMDA-Net effectively integrate features from multiple dimensions, resulting in improved classification performance. This success stems from integrating temporal convolutions and attention mechanisms, which effectively interpret EEG data. Designed to capture both temporal and spatial characteristics of EEG signals, deep learning classifiers have proven superior to traditional methods. The results of this study reveal a substantial enhancement in detection rates over existing models, highlighting the efficacy of the proposed approach for practical applications. The implications of this research are profound, extending beyond academic realms to inform the development of more sophisticated driver assistance systems. Incorporating this fatigue detection algorithm into these systems could significantly reduce fatigue-related incidents on the road, thus fostering safer driving conditions. This paper provides an exhaustive analysis of the dataset, methods employed, results obtained, and the potential real-world applications of the findings, aiming to contribute significantly to advancements in automotive safety.
\end{abstract}
\begin{keywords}
    Driver Fatigue | EEG | Safety | SEED-VIG | Vigilance
\end{keywords}

%%%%%%%%%%%%%%%%%%%%%%%%%%%%%%%%%%%%%%%%%%%%%%%%%%
%%%%%%%%%%%%%%%%%%%%%%%%%%%%%%%%%%%%%%%%%%%%%%%%%%

\begin{corrauthor}
%\texttt{r.henriques{@}ucl.ac.uk}
g\_siddhad\at cs.iitr.ac.in
\end{corrauthor}

%%%%%%%%%%%%%%%%%%%%%%%%%%%%%%%%%%%%%%%%%%%%%%%%%%
%%%%%%%%%%%%%%%%%%%%%%%%%%%%%%%%%%%%%%%%%%%%%%%%%%

\section{Introduction}
\label{sec_intro}

Enhancing road safety through effectively managing driver fatigue is paramount in the automotive industry, given its significant role in global road accidents. This prevalent issue impairs cognitive and motor functions, diminishing a driver's alertness and responsiveness to changing road conditions~\cite{delvigne2022spatio}. In light of these concerns, this study focuses on fatigue detection using advanced computational techniques applied to electroencephalogram (EEG) signals, a direct method has shown promise over traditional indirect methods such as monitoring steering wheel movements or analyzing eyelid closures.

Recent statistics indicate that driver fatigue is implicated in about 20\% of road accidents, underscoring the essential need for effective detection systems in modern vehicles~\cite{orru2021electroencephalography}. Unlike traditional approaches, which often result in delayed fatigue detection, EEG-based methods allow for real-time, accurate assessments by directly measuring neurological activity. These techniques utilize the distinct capabilities of EEG signals to mirror neurophysiological changes linked to fatigue, capturing specific brain wave patterns such as theta and alpha waves. This enables a precise evaluation of a driver's vigilance levels, which is unachievable through other methods~\cite{zuo2022driver}. Moreover, recent research has demonstrated the feasibility of decoding cognitive states such as attention and distraction in a real-life setting using EEG~\cite{kaushik2022decoding}. This suggests that EEG-based systems could potentially be used to identify a wider range of driver states, including those that may contribute to accidents beyond fatigue.

This study employs the SEED-VIG dataset~\cite{zheng2017multimodal}, renowned for its application in EEG-based vigilance estimation, facilitating the accurate examination of signals pertinent to real-world driving situations. The research enhances pattern recognition methods for robust feature extraction and effective classification of fatigue states by integrating traditional machine learning algorithms with deep neural networks. This dual approach significantly improves the accuracy and reliability of fatigue detection systems, effectively overcoming the constraints of existing models.

This paper introduces a unified lightweight NLMDA-Net to facilitate relevant feature extraction from complex EEG signals with the help of multi-dimensional attention modules. The contributions of this paper are as follows:
\begin{itemize}
    \item A lightweight network, NLMDA-Net, is proposed for driver fatigue detection using EEG data. It comprises the feature extraction capabilities of ConvNet and EEGNet.
    \item Channel Attention Module: The module leverages a tensor product to expand channel information into the depth dimension, enhancing the network's ability to process and analyze spatial features in EEG signals. This innovation increases sensitivity to spatial variations. Furthermore, the tanh function, a non-linear activation mechanism, stabilizes the learning process by normalizing amplitude variability. Its properties prevent the dying gradient problem and facilitate the capture of bi-directional relationships, which is essential for focusing the attention mechanism on the most informative EEG features.
    \item Parameter Efficiency: By reducing the number of convolution kernels as the network depth increases, NLMDA-Net tailors its architecture better to suit the predominant information-rich time domain of EEG signals, enhancing the network's efficiency and effectiveness.
    \item Adaptation to Data Scarcity: The network design is particularly suited for scenarios with limited EEG data, preventing over-fitting and accommodating EEG's low spatial resolution characteristics.
\end{itemize}

The structure of this paper is designed to methodically explore EEG-based fatigue detection and its implications for enhancing road safety technologies. Section~\ref{sec_related} reviews recent literature on driver drowsiness and vigilance. Section~\ref{sec_method} explains the methodology employed. Section~\ref{sec_result} presents the empirical findings. The paper is concluded in Section~\ref{sec_conclusion}, where the discussion extends to this research's implications and future directions.

%%%%%%%%%%%%%%%%%%%%%%%%%%%%%%%%%%%%%%%%%%%%%%%%%%
%%%%%%%%%%%%%%%%%%%%%%%%%%%%%%%%%%%%%%%%%%%%%%%%%%

\section{Related Work}
\label{sec_related}

Early research in EEG-based fatigue detection has primarily focused on identifying fatigue-associated biomarkers, such as the theta and alpha EEG frequency bands, such as variations in theta and alpha EEG frequency bands~\cite{huang2009tonic,jap2009using}. Subsequent advancements have introduced sophisticated signal processing techniques to improve detection accuracy, incorporating wavelet transforms and power spectral density analysis~\cite{ahmadi2021automated,chinara2021automatic}. The advent of deep learning has significantly transformed EEG analysis. In particular, Convolutional Neural Networks (CNNs) and Recurrent Neural Networks (RNNs) have become increasingly prevalent, appreciated for their adept handling of spatial and temporal data, respectively~\cite{sheykhivand2022developing,balam2021automated}.

Recent efforts have seen the development of hybrid models that combine CNNs with RNNs or other machine learning techniques to capitalize on their spatial and temporal feature extraction capabilities~\cite{ardabili2024novel,xu2021key}. Comparative studies of deep learning architectures indicate that CNNs provide superior accuracy and enhance computational efficiency, rendering them ideal for real-time applications~\cite{wang2022gradient}. Deep learning models generally surpass traditional machine learning methods due to their enhanced capacity to manage large, complex datasets without extensive feature engineering~\cite{wang2023recent,fouad2023robust}.

Furthermore, the application of transfer learning with pre-trained models on EEG data has demonstrated potential in mitigating the challenges posed by the need for large labelled datasets, which are often a limiting factor in EEG research~\cite{shalash2019driver}. Additionally, recent advancements in synthetic data generation~\cite{siddhad2024enhancing} also offer potential to augment real-world datasets and improve model performance. Transformers and attention mechanisms have emerged as powerful tools for EEG analysis~\cite{siddhad2024efficacy,kaushik2023motor}. Furthermore, recent studies have explored the integration of attention mechanisms into deep neural networks to enhance the identification of fatigue-related EEG features~\cite{gao2023multi,miao2023lmda}. However, challenges remain, such as the variability in EEG signals across individuals, which can affect model generalization~\cite{kar2010eeg}. Additionally, the presence of artifacts in EEG data due to head movements or external electrical interference continues to be a significant issue, potentially compromising the effectiveness of fatigue detection systems~\cite{kilicarslan2019characterization}.

%%%%%%%%%%%%%%%%%%%%%%%%%%%%%%%%%%%%%%%%%%%%%%%%%%
%%%%%%%%%%%%%%%%%%%%%%%%%%%%%%%%%%%%%%%%%%%%%%%%%%

\section{Methodology}
\label{sec_method}

\begin{figure*}[!t]
    \centering
    \includegraphics[width=\textwidth]{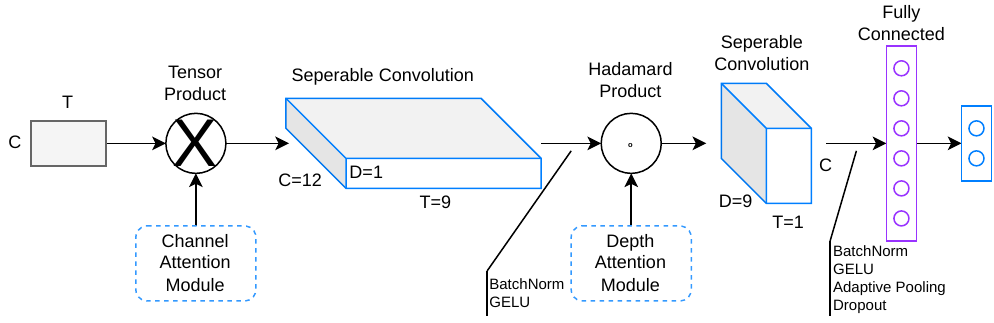}
    \caption{Architecture of NLMDA-Net: It comprises the benchmark network, the channel attention module, and the depth attention module. Channel attention is facilitated through a tensor product, and depth attention is achieved through a Hadamard product of tensors. NLMDA-Net's output varies with the number of task categories. Notably, C represents the number of EEG channels, and T signifies the time samples in a trial.}
    \label{fig_method}
\end{figure*}

The NLMDA-Net architecture shown in Figure \ref{fig_method} represents a novel integration of benchmark network capabilities enhanced by two specialized attention modules: the channel attention and the depth attention modules. The channel attention module is designed to strengthen the network's ability to discern relevant information within the spatial dimensions of EEG signals. Complementarily, the depth attention module aims to refine the representation of high-dimensional EEG features, ensuring a deeper and more targeted analysis. These modules are strategically developed to be compatible with any existing convolutional neural network structures.

As illustrated in Figure 1, the conceptual foundation of NLMDA-Net is intricately tied to the fundamental properties of EEG signals. Contrary to the prevailing trend in deep learning, which favours increasingly complex architectures~\cite{krizhevsky2017imagenet,simonyan2014very,szegedy2016rethinking,he2016deep}, insights from neuroscience~\cite{jensen2010shaping,pfurtscheller1999event,polich2007updating} advocate for the inherent simplicity of EEG characteristics. This simplicity suggests that even a shallower network architecture might suffice for effective EEG signal decoding.

Moreover, the typical scarcity of EEG data combined with the high data requirements of neural network models highlights the necessity for a more streamlined and lightweight network design. Such a design is essential to prevent over-fitting and accommodate the low spatial resolution of EEG and the diverse informational content across its temporal and spatial dimensions. The tailored approach provided by NLMDA-Net, depicted in Figure 1, is particularly suited to these unique challenges. 

The design considerations of NLMDA-Net and its compatibility with existing architectures are further detailed in the subsequent sections, providing a comprehensive overview of each component within the NLMDA-Net framework. This structured exposition underscores the architecture's potential to improve EEG-based applications through focused attention mechanisms and simplified network design.

%%%%%%%%%%%%%%%%%%%%%%%%%%%%%%%%%%%%%%%%%%%%%%%%%%

\subsection{Benchmark Network}
The NLMDA-Net architecture merges the foundational feature extraction capabilities of ConvNet~\cite{schirrmeister2017deep} with the advanced separable convolution technique from EEGNet~\cite{lawhern2018eegnet}, optimizing the extraction of temporal and spatial features from EEG signals. This architecture strategically employs a two-layer convolutional setup comprising a temporal convolutional layer and a spatial convolutional layer, utilizing separable convolutions to reduce the network's parameter count effectively.

In this approach, the temporal convolutional layer is characterized by a kernel size of (12, 1, 9), where 12 denotes the number of kernels, and the kernels' spatial and temporal dimensions are one and nine, respectively. Conversely, the spatial convolutional layer employs a kernel size of (7, C, 1), with $C$ representing the number of EEG channels. These consistent kernel dimensions are upheld throughout all experiments conducted within the NLMDA-Net framework, ensuring a standardized assessment of the network's efficacy in extracting features from EEG data.

Moreover, NLMDA-Net employs the Gaussian Error Linear Unit (GELU) activation function, as introduced by Hendrycks et al.~\cite{hendrycks2016gaussian}, offering improved smoothness compared to the Exponential Linear Unit (ELU)~\cite{clevert2015fast} utilized in previous models such as EEGNet and ConvNet. To effectively handle the typically substantial number of parameters required by fully connected layers, NLMDA-Net integrates adaptive average pooling. This technique dynamically adjusts the pooling kernel size to $(1, k_{pooling})$, where $k_{pooling}$ is detailed in Equation~\eqref{eq:1}. This adaptive approach ensures efficient parameter management while preserving the network's feature extraction capabilities.

\begin{equation}
    k_{pooling} = \text{max}(1, \lfloor f/10/N \rfloor)
    \label{eq:1}   
\end{equation}
where, the operator $\lfloor \rfloor$ represents the mathematical operation of rounding down to the nearest integer. It's applied concerning the input signal frequency, denoted by $f$, where $N$ signifies the number of training samples. The determination of $N$ is intricately tailored to accommodate the intricacies of EEG data collection, as elucidated by Equation~\eqref{eq:2}. This method ensures the parameters are finely tuned to align with the dataset's unique characteristics.

\begin{equation}
    N = \text{max}(1, \lfloor N_{t}/200\rfloor)
    \label{eq:2}   
\end{equation}
where, $N_{t}$ denotes the number of training samples, a pivotal parameter crucial for quantifying the dataset size utilized in model training. Its significance lies in its direct impact on the robustness and generalizability of the learned features.

NLMDA-Net distinctively adapts its architecture by reducing the number of convolution kernels from 12 to 1 as the network depth increases, a decision driven by two primary considerations. Firstly, employing a higher number of kernels in the spatial layers risks rapid over-fitting, leading to an exponential increase in the parameters of the fully connected layers, potentially compromising the network's capability to extract meaningful features. Secondly, considering that EEG signals predominantly contain richer information in the time domain than the spatial domain, it is pragmatic to allocate more kernels for extracting time domain features. This approach enhances the network's ability to effectively capture the most relevant data.

%%%%%%%%%%%%%%%%%%%%%%%%%%%%%%%%%%%%%%%%%%%%%%%%%%

\subsection{Channel Attention Module}
In EEG data acquisition, the signal captured by a single electrode channel is a composite of various neuronal activities influenced by volume conduction effects. Certain studies have employed source reconstruction techniques to enhance the spatial resolution of EEG signals to identify neuronal regions associated with specific EEG activities for in-depth analysis~\cite{ieracitano2020novel,cao2020psychological}. However, these techniques often require extensive prior knowledge and face integration challenges with end-to-end neural network models, complicating the decoding of EEG signals across different paradigms.

In the context of neural network architecture, models tailored explicitly for EEG decoding, such as EEGNet~\cite{lawhern2018eegnet}, ConvNet~\cite{schirrmeister2017deep}, and DRDA~\cite{zheng2022copula}, typically prioritize temporal convolutions over spatial ones. This approach can result in a relative neglect of spatial dimensions within EEG signals. To bridge this gap, a novel channel attention module is introduced that enhances the neural network's ability to assimilate spatial information from EEG data. This module draws conceptual parallels with source reconstruction techniques, acting on the input data to expand its spatial dimensions into the depth dimension through a Tensor product. This adaptation aims to improve the processing and analysis of spatial features within EEG signals, enhancing the overall efficacy of the neural network in decoding complex EEG data.

In this model, consider an EEG input sample denoted by $x$, where $x \in \mathbb{R}^{1*C*T}$. Here, $C$ represents the number of channels, and $T$ denotes the time samples. Additionally, a tensor $c$ is introduced, which follows a normal distribution, defined as $c \in\mathbb{R} ^{D*1*C}$, where $D$ corresponds to both the number of instances and convolutional kernels.

The channel attention module in the architecture utilizes a tensor product to project channel information from $x$ into the depth dimension, $D$. This operation preserves the spatial information inherent in the channel data and effectively integrates it with the following temporal convolution. The mathematical representation of this operation is outlined in the subsequent formula, illustrating how the module enhances the feature extraction capabilities of the neural network by augmenting the depth dimension with significant channel-specific information. This approach facilitates better analysis of EEG signals by leveraging both spatial and temporal dimensions efficiently. The operation is mathematically represented as:

\begin{equation}
    X_{hct}^{'}=\sum_{d} \textbf{X}_{dct}\textbf{C}_{hdc}
    \label{eq:3}   
\end{equation}

In the employed channel attention module, subscripts in the formula denote the respective dimensions, with matching subscript letters indicating that the two tensors share the same dimensionality in that specified dimension. This module introduces $D \times C$ trainable parameters, where $D$ is a hyper-parameter that can be optimized for specific tasks to enhance performance potentially. However, to maintain a consistent basis for comparison with NLMDA-Net and other benchmark models, $D$ is fixed at 9. This setting introduces significantly fewer parameters than traditional models and effectively maps spatial information into the depth dimension. This strategic approach offers a novel perspective on the attention mechanisms utilized for EEG signals.

The transformation of the input $X$ and its subsequent processing through the neural network's sequential components involves reshaping and applying linear layers. The initial step in this transformation process features a linear transformation followed by a non-linear activation, specifically using the $tanh$ function. This methodological choice facilitates the efficient integration and processing of EEG data, emphasizing the innovative use of attention mechanisms to enhance the depth dimension's role in spatial feature representation. The first linear transformation with a non-linear activation $tanh$ is represented as

\begin{equation}
    E = W_{2}*\tanh(W_{1}*X+b_{1})
    \label{eq:4}   
\end{equation}
\begin{equation}
    \alpha=\text{Softmax}(E)=\frac{\exp(E)}{\sum \exp(E)}
    \label{eq:5}   
\end{equation}
\begin{equation}
    \text{Context}_{b,c,n,t} = \alpha_{b,c,n} \cdot X_{b,c,t}
    \label{eq:6}
\end{equation}

%%%%%%%%%%%%%%%%%%%%%%%%%%%%%%%%%%%%%%%%%%%%%%%%%%

\subsection{Effect of Non-Linearity}
The hyperbolic tangent (tanh) function, renowned for its output range of [-1, 1], emerges as a potent normalization tool for processing EEG signals. These signals exhibit substantial amplitude fluctuations across diverse recording conditions and subjects. Normalizing such variations with tanh fosters stable learning dynamics, ensuring consistent neural network performance across heterogeneous datasets. Tanh's hallmark smoothness and continuous nature, coupled with a non-zero derivative across its operational span, are pivotal in facilitating gradient flow during backpropagation. This characteristic mitigates the risk of encountering the `dying gradient problem' prevalent in rectified linear units (ReLU), where gradients may diminish to zero, impeding further learning.

Furthermore, tanh's saturation at the extremities of its range offers resilience against outliers and extreme values in the data, facilitating more robust convergence during training. In contrast to linear activation functions like ReLU, tanh's capacity to yield positive and negative outputs enables the model to capture bi-directional data relationships effectively. This bi-directionality proves particularly advantageous in the context of the channel attention module within neural networks. Here, tanh synergizes with the softmax function, a staple in attention mechanisms, enhancing the latter's efficacy in spotlighting the most salient features in EEG data. The diverse output range of tanh empowers softmax to operate across a broad spectrum of values, thereby augmenting the attention mechanism's ability to emphasize informative data attributes.

%%%%%%%%%%%%%%%%%%%%%%%%%%%%%%%%%%%%%%%%%%%%%%%%%%

\subsection{Depth Attention Module}
In computer vision, feature maps in the depth dimension are often regarded as detectors of specific features within an input, identifying `what' is meaningful (\cite{woo2018cbam,zeiler2014visualizing}. This principle is crucial in models such as the Convolutional Block Attention Module (CBAM), where depth attention aggregates depth information through global pooling and dense layers to refine feature focus\cite{woo2018cbam}. However, this methodology proves less effective when decoding EEG signals due to the distinct nature of spatial and temporal dimensions in EEG data. In EEG, the global pooling and subsequent fully connected layers tend to oversimplify the depth information and drastically increase the model parameters, potentially leading to overfitting and degraded performance of the base network. A specialized depth attention module is proposed, tailored for EEG decoding to address these challenges. This module integrates concepts from local cross-depth interaction techniques, effectively balancing parameter efficiency and depth feature utilization. 

The depth attention module is strategically positioned between the temporal and spatial convolution layers, encompassing three main components: Semi-Global Pooling, Local Cross-Depth Interaction, and Adaptive Weighting. In contrast to conventional global pooling methods, Semi-Global Pooling averages the spatial dimensions while retaining temporal details, thus preserving a more comprehensive representation of depth features. Following this pooling, a convolutional layer is employed to encourage local interactions among features, substantially lowering the trainable parameters' count relative to fully connected layers. Subsequently, features undergo adaptive weighting and are transformed into probabilistic values via a softmax function. To maintain the amplitude sensitivity crucial for EEG signals, these softmax outputs are amplified to the level of the original inputs using a Hadamard product.
\begin{equation}
    M(F) = (\text{Softmax}\left(\text{Conv}\left(\text{Pooling}*(F)^T\right)\right)^{*} D^{'} )^{T}
    \label{eq:7}   
\end{equation}
where, $F \in \mathbb{R}^{D_o*C_o*T_o}$ represents the input feature tensor, capturing the dimensions of depth ($D_o$), channels ($C_o$), and temporal sequence ($T_o$) and $M(F) \in \mathbb{R}^{D_{o}*1*T_{o}}$ denotes the output feature map, simplifying the channels to one while maintaining depth and time dimensions. Pooling refers to the semi-global pooling operation, Conv indicates the convolution layer, and $T$ represents the transpose operation applied to the spatial and depth dimensions of the tensor.

%%%%%%%%%%%%%%%%%%%%%%%%%%%%%%%%%%%%%%%%%%%%%%%%%%
%%%%%%%%%%%%%%%%%%%%%%%%%%%%%%%%%%%%%%%%%%%%%%%%%%

\section{Results and Discussion}
\label{sec_result}

\subsection{Experimental Data}
The SEED-VIG dataset~\cite{zheng2017multimodal} is an open-source resource for investigating vigilance and driver drowsiness through EEG data collected from 23 participants to ensure diverse subject representation. Participants underwent a driving simulation resembling real-world conditions, enhancing the dataset's applicability for drowsiness studies. EEG recordings utilized 17 channels based on the 10–20 system, covering key temporal and posterior regions (FT7, FT8, T7, T8, TP7, TP8 for temporal; CP1, CP2, P1, PZ, P2, PO3, POZ, PO4, O1, OZ, O2 for posterior), ensuring comprehensive brain activity capture. Recorded at 1000Hz, the dataset offers high temporal resolution for detailed vigilance and drowsiness analysis. Fatigue induction was optimized by scheduling sessions post-lunch.

The drowsiness states are calculated as a percentage of eye closure time per unit time (PERCLOS). PERCLOS were categorized into `awake' and `drowsy' states at a 0.5 threshold. This binary classification enabled precise evaluation of this method's ability to detect driver fatigue. EEG signals are band-pass filtered between 1-75 Hz to reduce artifacts and down-sampled with a sampling frequency of 200 Hz. The dataset was epoched into one-second intervals, resulting in the shape of (1, channel count, EEG length), i.e., (1, 17, 200), yielding 40710 samples. The dataset is split into 70:15:15 ratios for train, validation, and test sets.

%%%%%%%%%%%%%%%%%%%%%%%%%%%%%%%%%%%%%%%%%%%%%%%%%%

\subsection{Implementation Details}
The experimental setup involved a DELL Precision 7820 Tower Workstation with Ubuntu 22.04 OS, Intel Core(TM) Xeon Silver 4216 CPU, and an NVIDIA RTX A2000 12GB GPU. This hardware facilitated the implementation of DL models using Python 3.10 and the PyTorch library. The Adam optimizer, known for its computational efficiency, was used with default parameters ($\eta$ = 0.001, $\beta_1$ = 0.9, $\beta_2$ = 0.999). EEGNet and TSception were trained for 100 epochs, with batches of 16 and a learning rate of $1e-4$. The Radial Basis Function (RBF) kernel from scikit-learn~\cite{sklearn} was used with default settings for SVM. Classification accuracy was determined through stratified five-fold cross-validation, averaging the results for comprehensive assessment.

%%%%%%%%%%%%%%%%%%%%%%%%%%%%%%%%%%%%%%%%%%%%%%%%%%

\subsection{Evaluation}

\begin{table}[!t]
    \centering
    \caption{Comparison of Classifier Performance for Detecting Driver Drowsiness Using SEED-VIG Dataset, Shown with 95\% Confidence Intervals}
    \label{tab_result}
    \renewcommand{\arraystretch}{1.1}
    \begin{tabular}{l c}
        \toprule
        \textbf{Classifier} & \textbf{Accuracy} \\
        \midrule
        \textbf{SVM}~\cite{cortes1995support} & $65.52 \pm 0.02$ \\
        \textbf{EEGNet}~\cite{lawhern2018eegnet} & $80.74 \pm 0.75$ \\
        \textbf{TSception}~\cite{ding2022tsception} & $83.15 \pm 0.36$ \\
        \textbf{ConvNext}~\cite{liu2022convnet} & $81.95 \pm 0.61$ \\
        \textbf{LMDA}~\cite{miao2023lmda} & $81.06 \pm 0.99$ \\
        \midrule
        \textbf{Proposed NLMDA-Net} & $\mathbf{83.71 \pm 0.30}$ \\
        \bottomrule
    \end{tabular}
\end{table}

\begin{figure}[!t]
    \centering
    \includegraphics[width=\linewidth]{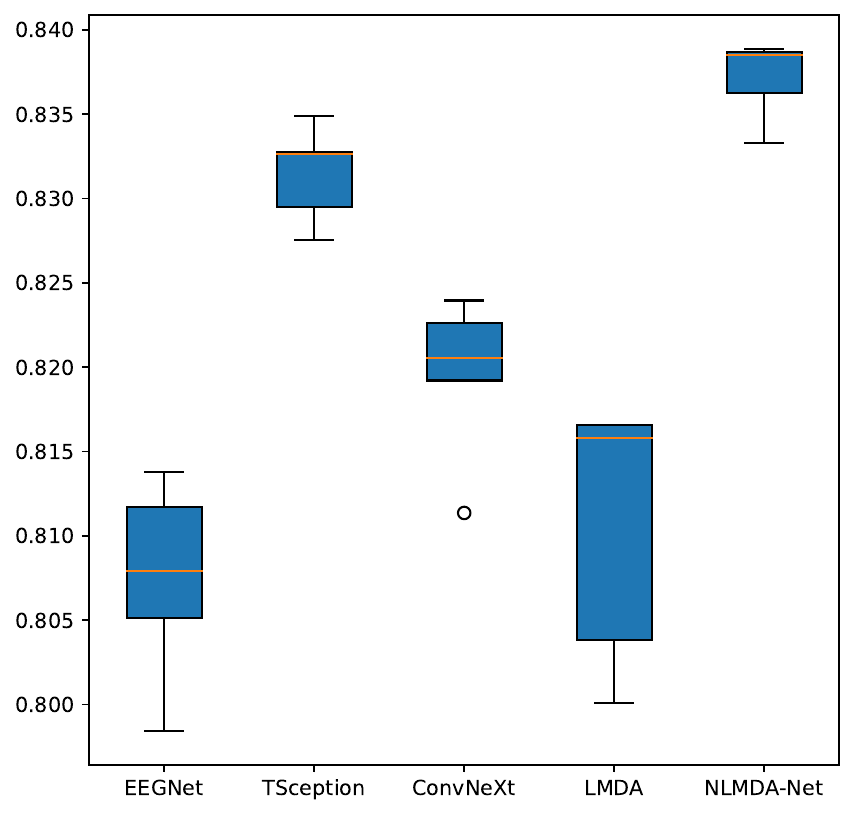}
    \caption{Boxplot Illustrating the Distribution of Classifier Accuracies for Driver Drowsiness Detection on the SEED-VIG Dataset} 
    \label{fig_boxplot}
\end{figure}

The data in Table~\ref{tab_result} compares various machine learning classifiers, analyzing their accuracy in detecting driver fatigue using the SEED-VIG dataset. This study encompasses a range of classifiers, each employing unique approaches and architectures designed to process and predict based on EEG data.

The SVM classifier exhibits the lowest accuracy at 65.52\%, suggesting its linear operational nature may be less effective at interpreting the complex patterns present in EEG signals, in contrast to more sophisticated, non-linear models. On the other hand, EEGNet, a neural network specifically optimized for EEG data processing, achieves an improved accuracy of 80.74\%. Its architecture, which adeptly handles both spatial and temporal dynamics of EEG signals, significantly outperforms traditional machine learning models like SVM.

Further analysis reveals that TSception and NLMDA-Net demonstrate the highest accuracies, with scores of 83.15\% and 83.71\%, respectively. These models incorporate advanced features such as temporal convolutions and attention mechanisms, enhancing their capability to capture subtle EEG signal changes associated with fatigue. ConvNext and LMDA also perform commendably, with accuracies of 81.95\% and 81.06\%, respectively. These classifiers benefit from recent advancements in convolutional network design and machine learning techniques tailored to handle large-scale, complex data structures typical of EEG datasets.

The results from Table~\ref{tab_result} and boxplot from Figure~\ref{fig_boxplot} consistently indicate that classifiers utilizing deep learning architectures, particularly those incorporating specialized mechanisms for extracting temporal and spatial features, surpass traditional machine learning methods in performance. This trend underscores the advantages of models that can adaptively learn from the intrinsic characteristics of EEG data related to drowsiness, suggesting a strategic direction for future development in this area. The confidence intervals reported also provide valuable insight into the consistency of each model's performance across different experimental setups, reinforcing the reliability of these findings.

%%%%%%%%%%%%%%%%%%%%%%%%%%%%%%%%%%%%%%%%%%%%%%%%%%
%%%%%%%%%%%%%%%%%%%%%%%%%%%%%%%%%%%%%%%%%%%%%%%%%%

\section{Conclusion}
\label{sec_conclusion}

This study addresses the critical issue of driver fatigue by applying advanced computational techniques to EEG signals. This provides a real-time, direct method for fatigue detection, surpassing traditional methods like monitoring steering movements or eyelid activity. Utilizing the SEED-VIG dataset, the research combines traditional machine learning and deep neural networks to refine pattern recognition techniques, enhancing the detection systems' accuracy and reliability. The results, methodology, and relevant literature are thoroughly explored, leading to discussions on the implications and future directions for enhancing road safety through improved fatigue detection technologies. Adopting deep learning, particularly CNNs and RNNs, has revolutionized EEG analysis by effectively handling spatial and temporal data, with hybrid models enhancing real-time feature extraction. Deep learning excels over traditional methods by managing large datasets with minimal feature engineering. Transfer learning and attention mechanisms have also emerged as solutions to challenges such as data variability and artifacts, improving signal quality and model generalizability. However, refining fatigue detection systems' accuracy and broad applicability remains a challenge. 

This study uses the SEED-VIG dataset to evaluate the efficacy of various machine-learning classifiers in detecting driver drowsiness. The results reveal that deep learning models, especially NLMDA-Net, show superior performance, achieving accuracy up to 83.71\%. These models excel due to their advanced features, such as temporal convolutions and attention mechanisms, effectively capturing EEG signal differences associated with fatigue. This suggests a significant potential for deep learning approaches to enhance fatigue detection systems, advocating for a strategic pivot towards these technologies to improve the accuracy and generalizability of drowsiness detection methods. The consistency of model performances, supported by confidence intervals, reinforces the reliability of these findings. Future research in EEG-based fatigue detection should prioritize advancements that bolster accuracy and usability. Integrating multimodal data, including heart rate variability, eye tracking, and contextual driving information, can enrich the understanding of the driver's state, facilitating a more comprehensive analysis.

%%%%%%%%%%%%%%%%%%%%%%%%%%%%%%%%%%%%%%%%%%%%%%%%%%
%%%%%%%%%%%%%%%%%%%%%%%%%%%%%%%%%%%%%%%%%%%%%%%%%%

% \begin{acknowledgements}
% \blindtext
% \end{acknowledgements}

%%%%%%%%%%%%%%%%%%%%%%%%%%%%%%%%%%%%%%%%%%%%%%%%%%
%%%%%%%%%%%%%%%%%%%%%%%%%%%%%%%%%%%%%%%%%%%%%%%%%%

\section*{Bibliography}
\bibliography{references}

%%%%%%%%%%%%%%%%%%%%%%%%%%%%%%%%%%%%%%%%%%%%%%%%%%
%%%%%%%%%%%%%%%%%%%%%%%%%%%%%%%%%%%%%%%%%%%%%%%%%%

%% You can use these special %TC: tags to ignore certain parts of the text.
%TC:ignore
%the command above ignores this section for word count
% \onecolumn
% \newpage

% \section*{Word Counts}
% This section is \textit{not} included in the word count. 
% \subsection*{Notes on Nature Methods Brief Communication}
% \begin{itemize}
% \item Abstract: 3 sentences, 70 words.
% \item Main text: 3 pages, 2 figures, 1000-1500 words, more figures possible if under 3 pages
% \end{itemize}

% \subsection*{Statistics on word count}
% \detailtexcount
% \newpage

%%%%%%%%%%%%%%%%%%%%%%%%%%%%%
% Supplementary Information %
%%%%%%%%%%%%%%%%%%%%%%%%%%%%%
% \captionsetup*{format=largeformat}
% \section{Something about something} \label{note:Note1} 
% \Blindtext

%TC:endignore
%the command above ignores this section for word count

%%%%%%%%%%%%%%%%%%%%%%%%%%%%%%%%%%%%%%%%%%%%%%%%%%
%%%%%%%%%%%%%%%%%%%%%%%%%%%%%%%%%%%%%%%%%%%%%%%%%%

\end{document}